\documentstyle[sprocl]{article}

\def\Journal#1#2#3#4{{#1} {\bf #2}, #3 (#4)}

\def\NPB{{\em Nucl. Phys.} B}
\def\PLB{{\em Phys. Lett.}  B}

\def\PRD{{\em Phys. Rev.} D}
\def\EPJC{{\em Eur. Phys. J.} C}
\def\IJMPA{{\em Int. J. Mod. Phys.} A}
\def\PRP{\em Phys. Rep.}
\def\PTP{{\em Prog. Theo. Phys.} (PS)}

\def\be{\begin{equation}}
\def\ee{\end{equation}}
\def\bea{\begin{eqnarray}}
\def\eea{\end{eqnarray}}
\def\gm{\Gamma}
\def\ln{\left<}
\def\rn{\right>}
\def\be{\begin{equation}}
\def\ee{\end{equation}}
\def\bea{\begin{eqnarray}}
\def\eea{\end{eqnarray}}
\def\gm{\Gamma}
\def\al{\alpha_s}
\def\ln{\left<}
\def\rn{\right>}
\def\lln{\left(}
\def\rln{\right)}
\def\lq{\Lambda_{QCD}}
\def\bl{\bar \Lambda}

\def\ra{\rightarrow}

\def\om{\omega}
\def\qq{\ln \bar q q \rn}
\def\lbb{\Lambda_b}
\begin{document}

\title{PATTERN OF LIFETIMES OF BEAUTY HADRONS AND QUARK-HADRON
DUALITY IN HEAVY QUARK EXPANSION\footnote{Talk given at IV Workshop on
Continuous Advances in QCD, at Theoretical Physics Institute, University of
Minnesota, Minneapolis, USA during May 12-14, 2000}}

\author{S. ARUNAGIRI}

\address{Department of Nuclear Physics, University of Madras,\\
Guindy Campus, Chennai 600 025, Tamil Nadu, INDIA}

\maketitle\abstracts{
We discuss 
(i) the evaluation of the expectation values of four-quark
operators assuming that the heavy quark expansion for $b$ sector
converges at the third order in $1/m_Q$, and
(ii) the estimation of the duality breaking short distance nonperturbative corrections
to the parton decay rate.
We finally point out the implications of the result obtained for the assumption
of quark-hadron duality in heavy quark expansion.}
\vskip0.5cm
\section{Introduction}
In the heavy quark limit, the inclusive decay rate of weakly decaying heavy 
hadron ($H_Q$) can be expressed in an expansion of the weak matrix elements in 
powers of $1/m_Q$, where $m_Q$ is the heavy quark mass \cite{bigi}. (In this
article, though
we discuss generally for a heavy system, $b$ or $c$ but quantitatively refer 
only to beauty case). At the leading order,
the hadronic decay rate is given by the free heavy quark decay rate, universal
for all hadrons of given flavour quantum number, $Q$:
\be
	\Gamma_0 = {G_f^2 |V_{KM}|^2 m_Q^5 \over {192 \pi^3}}
\ee
where $|V_{KM}|$ is the relevant CKM matrix element. The leading order receives
corrections due to the motion of the heavy quark inside the hadron ($\lambda_1$)
and the heavy quark spin projection ($\lambda_2$), which appear at $1/m_Q^2$ in 
the expansion:
\bea
	\lambda_1 = \mu_\pi^2(H) = {1 \over {2M_H}}
		\left<H|\bar Q (iD)^2 Q|H\right> \\
	\lambda_2 = \mu_G^2(H) = {1 \over {2M_H}}
		\left<H|\bar Q g \sigma G Q|H\right> 
\eea
Numerically $\lambda_1(B)$ = -- 0.4 $GeV^2$, 
$\lambda_1(\Lambda_b)$ = -- 0.3 $GeV^2$
and $\lambda_2(B)$ = 0.12 $GeV^2$. The term $\lambda_2$ vanishes
for all baryon except $\Omega_Q$. Thus at $O(1/m_b^2)$, the total decay rate is
the sum of the above terms:
\be
	\Gamma(H_b) = \Gamma_0 \left\{1-{\lambda_1 - 3\lambda_2 \over {2m_b^2}}
		+ {2 \lambda_2 \over {m_b^2}} + ...\right\}
\ee
Thus, at $O(1/m_b^2)$, the decay rate splits up into mesonic and baryonic ones. 
The decay 
rate is the same for all $b$ mesons and for all $b$ baryons (except 
$\Omega_Q$). At this order, the theoretical predictions for the ratio of
lifetimes of $B$ mesons agree with the corresponding experimental values,
but not the ratio of lifetimes of $\Lambda_b$ baryon and $B$ meson \cite{caso}.
\vskip0.2cm
\begin{tabular}{c|c|c}
\hline
				& Theory, at $O(1/m_b^2)$ & experiment \\	
\hline
$\tau(B^-)/\tau(B^0)$		& 1			  & 1.04 $\pm$ 0.04 \\
$\tau(B^0_s)/\tau(B^0)$		& 1 $\pm$ 0.01		  & 0.99 $\pm$ 0.05 \\
$\tau(\Lambda_b)/\tau(B^0)$	& 0.9			  & 0.79 $\pm$ 0.06 \\
\hline
\end{tabular}
\vskip0.3cm
The discrepancy in the case of $\tau(\Lambda_b)/\tau(B^0)$ signifies something
interesting beyond the second order. The terms at $O(1/m_Q^3)$ are really
interesting because evaluation of the matrix elements  elude a concrete understanding of evaluation. Also
the discrepancy gives rise to doubt the validity of the underlying assumption
of the heavy quark expansion, {\it quark-hadron duality}. The third order in
$1/m_Q$ terms are involving both heavy and light quark fields. 
\be
	C(\mu){1 \over {2M_H}}
		\left<H|(\bar Q \Gamma_\mu q)(\bar q \Gamma^\mu Q|H\right>
\ee
where the Wilson coefficients, $C(\mu)$, describe the light quark processes
such as Pauli interference, Weak annihilation and $W$ scattering and the term
$\left<.....\right>$ is, in traditional terms, understood as probability of 
finding both the heavy and light quarks at zero separation, in other words 
the wavefunction density at origin, $|\Psi(0)|^2$. The evaluation of the
operators depends upon their parameterisation. 

In this talk, I discuss the evaluation of the expectation values of four-
quark operators from the difference in total decay rates (Sec. 2). It is based on the
assumption that at $O(1/m_b^3)$, the heavy quark expansion converges. The
short distance nonperturbative corrections are given in Sec. 3. This
signifies that the quark-hadron duality holds good within a few percent
of uncertainty due to {\it actual violation} (Sec. 4).

\section{Splitting of Total Decay Rates}
The estimation of expectation values of four-quark operators (EVFQO) is
as follows \cite{arun0}.
Unlike in the charmed sector, the HQE works well for beauty case
because of $m_b$ being asymptotically heavy. The spectator effects coming
from the term at $O(1/m^3)$ (hereinafter $m$ refers to the mass of $b$ quark),
is smaller but yet significant. If at all there exists any contribution
due to terms beyond $O(1/m^3)$, then it should be too small to be ignored.
Hence, being an asymptotic expansion, the HQE can be assumed to be convergent
at $O(1/m^3)$. This assumption does not hold for charm, since
\be
{16 \pi^2 \over {m_c^3}} C(\mu) \left< O_6 \right>_{H_c}   \gg
{16 \pi^2 \over {m_b^3}} C(\mu) \left< O_6 \right>_{H_b}   
\ee
where $C(\mu)$ stands for some structure involving $c_\pm$ and 
$\left< O_6 \right>_H$ the dimension six FQO of hadron.
In this background, we make use of the decay rates to obtain the 
EVFQO.  

The $B$ mesons, $B^-$, $B^0$ and $B^0_s$, are triplet
under $SU(3)_f$ flavour symmetry. Their total decay rate
splits up due to its light quark flavour dependence at the third order
in the HQE. The differences in the decay rates of the triplet,
$\gm(B^0) - \gm(B^-), \gm(B^0_s) - \gm(B^-)$ and
$\gm(B_s^0) - \gm(B^0)$, are related to the third order terms
in $1/m$ by
\bea
d\Gamma_{B^0-B^-} &=& -\Gamma^\prime_0 (1-x)^2\nonumber\\
&& \times \left\{Z_1{1 \over 3}(c_0+6)+(c_0+2)\right\}
\left<O_6\right>_{B^0-B^-} \label{one}\\
d\Gamma_{B^0_s-B^-} &=& -\Gamma^\prime_0 (1-x)^2\nonumber\\
&& \times \left\{Z_2{1 \over 3}(c_0+6)+(c_0+2)\right\}
\left<O_6\right>_{B^0_s-B^-}\\
d\Gamma_{B^0_s-B^0} &=& -\Gamma^\prime_0 (1-x)^2\nonumber\\
&& \times \left\{(Z_1-Z_2){1 \over 3}(c_0+6)\right\}
\left<O_6\right>_{B^0_s-B^0} \label{three}
\eea
where $d\gm_{B^0 - B^-} = d\gm(B^0) -\gm(B^-),
\gm^\prime_0 = 2G_f^2|V_{cb}|^2m_b^2/3\pi$,
$c_0$ = $2c_+-c_-$, $x$ = $m_c^2/m_b^2$ and
\bea
Z_1 &=& \left(cos^2\theta_c(1+{x \over 2})+
sin^2\theta_c \sqrt{1-4x}(1-x)\right)\\
Z_2 &=& \left(sin^2\theta_c(1+{x \over 2})+
cos^2\theta_c \sqrt{1-4x}(1-x)\right)\\
\left<O_6\right> &\equiv&
\left<{1 \over 2}(\bar b \Gamma_{\mu} b)[(\bar d \Gamma_{\mu} d)
-(\bar u \Gamma_{\mu} u)]\right> = 
\left<{1 \over 2}(\bar b \Gamma_{\mu} b)[(\bar s \Gamma_{\mu} s)
-(\bar q \Gamma_{\mu} q)]\right>
\eea
 with $q = u, d$.
In eqns. (\ref{one}-\ref{three}), 
the $rhs$ contains the terms corresponding to the
unsuppresed and suppressed nonleptonic decay rates and
twice the semileptonic decay rates at the third order.

On the other hand, for the triplet baryons,
$\Lambda_b$, $\Xi^-$ and $\Xi^0$, with
$\tau(\Lambda_b)$ $<$ $\tau(\Xi^0) \approx \tau(\Xi^-)$,
we have the relation between
the difference in the total decay rates \cite{vol} and the terms of
$O(1/m^3)$ in the HQE, as
\be
d\Gamma_{\Lambda_b-\Xi^0} = {3 \over 8}\Gamma_0^\prime (c_{00}-2)
\left<O_6\right>_{\Lambda_b-\Xi^0}
\ee
where $c_{00}$ = $-c_+(2c_-+c_+)$. 

\subsection{$\tau(\lbb)/\tau(B)$}
For the decay rates \cite{caso} $\gm(B^-) = 0.617 ps^{-1},     
\gm(B^0) = 0.637 ps^{-1}$ and $\gm(B^0_s) = 0.645 ps^{-1}$,
the EV$_{FQO}$ are obtained for $B$ meson, as an average from
eqs. (\ref{one}-\ref{three}):
\be
\left<O_6\right>_B = 8.08 \times 10^{-3} GeV^3.
\label{bval}
\ee
This is smaller than the one obtained
in terms of the leptonic decay constant, $f_B$.

We obtain the EV$_{FQO}$ for
the baryon
\be
\left<O_6\right>_{\Lambda_b-\Xi^0} = 3.072 \times 10^{-2} GeV^3
\label{lval}
\ee
where we have used the
decay rates corresponding to the lifetimes 1.24 ps and 1.39 ps 
of $\Lambda_b$ and $\Xi^0$ respectively \cite{caso}.
The EV$_{FQO}$ 
for baryon is about 3.8 times larger than that of B. For
these values
\be
\tau(\Lambda_b)/\tau(B) = 0.78
\ee
Using the experimental value of $\tau(B^-)$ = 1.55 ps
alongwith the above theoretical value,
the lifetime of $\Lambda_b$ turns out to be
\be
\tau(\Lambda_b) = {\Gamma(\Lambda_b) \over {\Gamma(B)}} \tau(B^-) =
1.20~ps.
\ee
The predictions here are significant qualitatively. This value
may change by few percent due to uncertainties. It is due to structure
of currents involved.

\section{Renormalon Corrections}
\subsection{Power Corrections} 
In this section, we present a study \cite{arun00} on 
the renormalons corrections considering the
heavy-light correlator in the QCD sum rules approach, assuming that the
nonperturbative short distance corrections given by the gluon mass that is much
larger than the QCD scale. We carry out the analysis for both heavy meson
and heavy baryon. Our study shows that the short distance
nonperturbative corrections to the baryon and the meson
differ by a small amount which is significant for the smaller 
lifetime of the $\Lambda_b$.

Let us consider the
correlator of hadronic currents $J$:
\be
\Pi(Q^2) = i \int d^4x e^{iqx} \ln 0|T\{J(x)J(0)\}|0\rn
\ee
where $Q^2 = -q^2$. The standard OPE is expressed as
\be
\Pi(Q^2) \approx [~\mbox{parton model}](1+a_1 \al + a_2 \al^2 + ....)
+ O(1/Q^4)
\ee
where the power suppressed terms are quark and gluon operators.
The perturbative series in the above equation can be rewritten as
\be
D(\al) = 1 + a_0 \al + \sum_{n = 1}^\infty a_n \al^n \label{d}
\ee
where the term in the sum is considered to be the
nonperturbative short distance quantity.
It is studied by Chetyrkin {et al} \cite{chet} assuming that the short
distance tachyonic gluon mass, $\lambda^2$,
imitates the nonperturbative physics of the QCD. This, for the gluon
propagator, means:
\be
D_{\mu \nu}(k^2) = {\delta_{\mu \nu} \over {k^2}} \ra
\delta_{\mu \mu}\left( {1 \over {k^2}}+{\lambda^2 \over {k^4}}\right)
\ee
The nonperturbative short distance corrections are argued to be the
$1/Q^2$ correction in the OPE.

Let us consider the assumption of
the gluon mass $\lambda^2 \gg \lq^2$ which
is not necessarily to be tachyonic one.
The feature of the assumption can be seen
with the heavy quark potential
\be
V(r) = -{4\alpha(r) \over {3r}}+kr \label{v}
\ee
where $k \approx$ 0.2 GeV$^2$,
representing the string tension. It has been argued in
\cite{bal} that the linear term can be replaced by a term of order $r^2$.
It is equivalent to replace $k$ by a term describing the ultraviolet
region.
For the potential in (\ref{v}),
\be
k \ra constant \times \al \lambda^2
\ee
In replacing the coefficient of the term of $O(r)$ by $\lambda^2$,
we make it consistent by the renormalisation factor.
Thus the coefficient $\sigma(\lambda^2)$ is given by \cite{ani}:
\be
\sigma(\lambda^2) = \sigma(k^2)\left(\alpha(\lambda^2) \over
\alpha(k^2)\right)^{18/11} \label{rge}
\ee
Introduction of $\lambda^2$ brings in a small correction to the Coulombic term.
By use of (\ref{rge}), we specify the effect at both the ultraviolet
region
and the region characterised by the QCD scale.
Then, we rewrite (\ref{d}) as
\be
D(\al) = 1 + a_0 \al\left(1+{k^2 \over {\tau^2}}\right)
\label{dc}
\ee
where $\tau$ is some scale relevant to the problem and $k^2$ should be
read
from (\ref{rge}). We would apply this to
the heavy light correlator in heavy quark effective theory.

We should note that in the QCD sum rules approach, the scale involved
in is given by the Borel variable which is about 0.5 GeV. But
in the heavy quark expansion the relevant scale is
the heavy quark mass, greater than the
hadronic scale. Thus, there it turns out to be infrared renormalons
effects.
But, still it represents the short distance nonperturbative
property, by virtue of the gluon mass being as high as the hadronic
scale.

\subsection{Meson} 
For the heavy light current, $J(x) = \bar Q(x) i\gamma_5
q(x)$,
the QCD sum rules is already known \cite{dai}:
\be
{\tilde f}_B^2 e^{-(\bl)} = {3 \over {\pi^2}}
\int_0^{\om_c} d \om \om^2 e^{-\om/\tau}D(\al) - \qq +
{1 \over {16 \tau^2}}\ln g \bar q \sigma G q \rn +...\label{hl}
\ee
where $\om_c$ is the duality interval, $\tau$ the Borel variable and
$D(\al)$ as defined in (\ref{d}), but of the form defined in (\ref{dc}).
It is, corresponding to the particular problem of heavy quarks, given as:
\be
D(\al)_B = 1 + {a_B \al}\left[1+{\lambda^2 \over {\tau^2}}
\left({\alpha(\lambda^2) \over {\alpha(\tau^2)}}\right)^{-18/11}\right]
\ee
where $a_B =   17/3+4\pi^2/9-4$log$(\om/\mu)$, with $\mu$
is chosen to be 1.3 GeV.

With the duality interval of about 1.2-1.4 GeV which is little smaller than
the
onset of QCD which corresponds to 2 GeV and $\bl \geq$ 0.6 GeV,
we get 
\be
\lambda^2 \approx 0.35 GeV^2. \label{l2m}
\ee

\subsection{Baryon}
For the heavy baryon current
\be
j(x) = \epsilon^{abc}(\bar q_1(x)C \gamma_5 t q_2(x))Q(x)
\ee
where $C$ is charge conjugate matrix, $t$ the antisymmetric flavour matrix
and $a, b, c$ the colour indices, the QCD sum rules is given \cite{dai} by
\bea
{1 \over 2} f_{\Lambda_b}^2 e^{\bl/\tau} &=&
{1 \over {20\pi^4}}\int_0^{\om_c} d \om \om^5
e^{-w/\tau}D(\al)_{\lbb}\\
&&+
{6 \over {\pi^4}}E_G^4 \int_0^{\om_c} d \om e^{-\om/\tau}+
{6 \over {\pi^4}}E_Q^6e^{-m_0^2/8\tau^2}
\eea
where 
\be
D(\al)_{\lbb} = 1-{\al \over {4\pi}}a_{\lbb} \left(1+{\lambda^2 \over {\tau^2}}\right)
\ee
with $\eta^\prime = r_1$log$(2\om/\mu)-r2)$. With
$f_{\Lambda_b}^2$ = $0.2 \times 10^{-3}$~GeV$^6$,
$\ln \bar q q \rn  = -0.24^3$ GeV$^3$, $\ln g \bar q \sigma G q \rn =
m_0^2 \ln \bar q q \rn$, $m_0^2 = 0.8$ GeV$^2$, $\ln \al GG \rn
= 0.04$ GeV$^4$ and $D(\al)_{\Lambda_b}$ is expressed in accordance with
power correction factor found in \cite{dai}.
As in the meson case, we obtain 
\be
\lambda^2 \approx 0.4 GeV^2.
\ee
Now we turn to the heavy quark expansion. The total decay
rate of a weakly decaying heavy hadron is, at the leading order, given by
\be
\Gamma(H_b) = \Gamma_0
\left[1-{\al \over \pi}\left( {2 \over 3}g(x) - \xi \right) \right] \label{rt}
\ee
where 
\be
\Gamma_0 = {G_f^2 |V_{KM}|^2 m_b^5 \over {192 \pi^3}} f(x)
\ee
As already mentioned, the power corrections are given by
the IR renormalons:
\bea
D_{IR} &=& \tilde a \al \lln 1+\xi \rln\\
\xi &=&
{\lambda^2 \over {m_b^2}} \lln {\alpha(\lambda^2)
\over {\alpha(m_b^2)}} \rln ^{11/18} 
\eea
In (\ref{rt}), the factor $\xi$ corresponds to the IR renormalons
which corresponds to the square root of the $\lambda^2$ term in the
above equation. These corrections are estimated to be 0.2$\Gamma_0$
and 0.21$\Gamma_0$
for $B$ and $\Lambda_b$ respectively. This is significant in view
of the discrepancy between the lifetimes of $B$ and $\Lambda_b$
being 0.2 ps$^{-1}$ with $\Gamma(B)$ = 0.68 ps$^{-1}$ and
$\Gamma(\Lambda_b)$ = 0.85 ps$^{-1}$.

\section{Quark-Hadron Duality}
The notion of quark-hadron duality (or duality)
is variably defined in the literature depending upon the problem at hand.
In the case of HQE description of inclusive decays too, there are 
many definitions.
However, the popular and simple formulation is that the hadronic quantities 
can be obtained in terms of quarks and gluons. As regards the heavy 
hadron decays, the idea of duality is that the sum of the exclusive decay rates
are given by sufficiently inclusive decay rate. It
is difficult to provide an analytical or a semi-analytical prescription
to establish duality. 

We are able to evaluate the EVFQO in a model independent way. Our results for 
the ratio $\tau(\lbb)/\tau(B)$ is closer to that of the experimental values.
The basic idea is the assumption on the convergence of the HQE for beauty case
at the third order in $1/m$. If we consider that the lifetime of hadron predicted by HQE
can be used as a validation test of duality, then we are led to believe that
duality holds good. If there is any discrepancy, it should be few percent and the same
can be neglected. 

On the other hand, the renormalon contribution to
the parton decay rate is found to be less than 10\%. This corrections
can be construed as duality breaking one. This again validates the assumption
of duality. Note that the renormalon corrections here are nonperturbative nature.

\section{Final Remarks}
Studies show that the the theoretical and experimental discrepancy on the ratio
of lifetimes of $\lbb$ and $B$ is due to the spectator effects. The basic issue
is the estimation of EVFQO. Previous studies leads to partial explanation of the discrepancy
\cite{rosner,neubert,colangelo}. However, in \cite{arun0}. a model independent evaluation has been made
which predicts the ratio, $\tau(\lbb)/\tau(B)$, close to the experimental value.
It is found in
agreement with the one obtained in a crude potential model \cite{arun}.
Also a QCD sum rules
approach predicted the ratio closer to the experimental value \cite{liu}.

Further, the notion of duality holds good in the HQE for the beauty case within
an amenable few percent. The same cannot be true for the charmed sector.

\section*{Acknowledgments}
I thank Prof. M. Voloshin for invitation to the workshop, for local hospitality
and useful discussions. 
Discussions with Prof. P. R. Subramanian is acknowledged.

\end{document}